\documentclass[conference]{IEEEtran}
\usepackage[T1]{fontenc}
\usepackage{amsmath,amssymb,amsfonts}
\usepackage{graphicx}   
\usepackage{textcomp}
\usepackage{xcolor}
\usepackage{float}
\usepackage{algorithm}
\usepackage{algpseudocode}
\usepackage{hyperref}
\usepackage{subcaption}
\usepackage[style=ieee, maxbibnames=6]{biblatex}
\graphicspath{{./images}}
\def\BibTeX{{\rm B\kern-.05em{\sc i\kern-.025em b}\kern-.08em
    T\kern-.1667em\lower.7ex\hbox{E}\kern-.125emX}}
    
\DefineBibliographyStrings{english}{%
    urlseen = {Accessed},
}

\usepackage{listings}
\usepackage{color}
\usepackage[justification=centering]{caption}

\usepackage{ifthen}

\makeatletter
\newcounter{IEEE@bibentries}
\renewcommand\IEEEtriggeratref[1]{%
  \renewbibmacro{finentry}{%
    \stepcounter{IEEE@bibentries}%
    \ifthenelse{\equal{\value{IEEE@bibentries}}{#1}}
    {\finentry\@IEEEtriggercmd}
    {\finentry}%
  }%
}
\makeatother

\definecolor{dkgreen}{rgb}{0,0.6,0}
\definecolor{gray}{rgb}{0.5,0.5,0.5}
\definecolor{mauve}{rgb}{0.58,0,0.82}

\begin{document}

\title{Serverless Approach to Sensitivity Analysis of Computational Models\\

}

\author{\IEEEauthorblockN{
Piotr Kica\IEEEauthorrefmark{1}\IEEEauthorrefmark{4},
Magdalena Otta\IEEEauthorrefmark{1}\IEEEauthorrefmark{2}\IEEEauthorrefmark{3},
Krzysztof Czechowicz\IEEEauthorrefmark{2}\IEEEauthorrefmark{3},
Karol Zając\IEEEauthorrefmark{1}\IEEEauthorrefmark{4},\\
Piotr Nowakowski\IEEEauthorrefmark{1},
Andrew Narracott\IEEEauthorrefmark{2}\IEEEauthorrefmark{3},
Ian Halliday\IEEEauthorrefmark{2}\IEEEauthorrefmark{3},
Maciej Malawski\IEEEauthorrefmark{1}\IEEEauthorrefmark{4}
}
\IEEEauthorblockA{\IEEEauthorrefmark{1}Sano Centre for Computational Medicine, Krak\'ow, Poland (\url{https://sano.science/})\\
}
\IEEEauthorblockA{\IEEEauthorrefmark{2}
Department of Infection, Immunity and Cardiovascular Disease, University of Sheffield, Sheffield, UK\\
}
\IEEEauthorblockA{\IEEEauthorrefmark{3}
Insigneo Institute for \textit{in silico} Medicine, University of Sheffield, Sheffield, UK\\
}
\IEEEauthorblockA{\IEEEauthorrefmark{4}Institute of Computer Science,
AGH University of Science and Technology, Krak\'ow, Poland\\
}
}

\maketitle

\begin{abstract}
Digital twins are virtual representations of physical objects or systems used for the purpose of analysis, most often via computer simulations, in many engineering and scientific disciplines. Recently, this approach has been introduced to computational medicine, within the concept of Digital Twin in Healthcare (DTH). Such research requires verification and validation of its models, as well as the corresponding sensitivity analysis and uncertainty quantification (VVUQ).

From the computing perspective, VVUQ is a computationally intensive process, as it requires numerous runs with variations of input parameters. Researchers often use high-performance computing (HPC) solutions to run VVUQ studies where the number of parameter combinations can easily reach tens of thousands. However, there is a viable alternative to HPC for a substantial subset of computational models - serverless computing.

In this paper we hypothesize that using the serverless computing model can be a practical and efficient approach to selected cases of running VVUQ calculations. We show this on the example of the EasyVVUQ library, which we extend by providing support for many serverless services. The resulting library - CloudVVUQ - is evaluated using two real-world applications from the computational medicine domain adapted for serverless execution. Our experiments demonstrate the scalability of the proposed approach.

\end{abstract}


\section{Introduction}

Digital twins are virtual representations of physical objects or systems, which can be used to accurately represent their counterparts for the purpose of analysis, most often via computer simulations. Digital twins have been used in many engineering and scientific disciplines, such as civil engineering, factory design, manufacturing or environmental studies. Recently, this approach has been introduced to computational medicine, within the concept of Digital Twin in Healthcare. A special case or often a crucial component of a digital twin is a computational model which we will focus on in this paper.

In computational medicine, modeling and simulation are becoming increasingly important, with a wide range of methods being applied to various physiological and pathological conditions. Examples include physics-based simulations analogous to those used in engineering such as computational fluid dynamics (CFD) for blood flow simulation \cite{Morrisheartjnl-2015-308044} \cite{Morris_Coronary_Blood_Flow}, or finite element method (FEM) modeling for computing the likelihood of bone fracture \cite{BENEMERITO2021106200}. Agent-based modeling can be used to simulate interactions in the immune system \cite{UISS-TB}, while multiscale modeling is useful in predicting the growth of tumors \cite{tumours_growth}. Other simplified or surrogate models have been developed to reduce the computational cost of such simulations, including 0-D and 1-D models to simulate haemodynamics in the cardiovascular system \cite{0D-1D_blood_flow}, reduced-order modeling techniques or machine learning (ML) models, which are typically inspired by physics-based solutions \cite{Benemerito2022}.

The multitude of successful applications of digital twins in healthcare research, along with medical modeling and simulation techniques, has led to their potential application in clinical decision support systems or digital therapeutics solutions. In order to bring these solutions to the market, a complex regulatory procedure needs to be followed through agencies such as the FDA in the US, or the EMA in Europe. One important stage in the process is performing verification and validation of the models, along with the corresponding sensitivity analysis and uncertainty quantification \cite{HUBERTS201868}. These procedures are subject to standards, such as ASME VV40 \cite{VV40}. 

From the computing perspective, VVUQ is a computationally intensive process, as it requires numerous runs of the model with variations of input parameters. Such studies are examples of high-throughput computing, involving many independent tasks of a similar nature but with different input data or parameter variations. Researchers typically use high-performance computing machines for running VVUQ studies, especially when the models are large-scale parallel computing simulations, as in the case of 3-D models using CFD or FEM methods, which often require several hours of computing time on multiple nodes of a cluster. The number of parameter combinations and therefore the number of tasks can easily reach tens of thousands, or more.

There are cases, however, when it would be reasonable to look for other computing infrastructures to solve VVUQ problems. One involves users who may not have access to publicly funded research infrastructures, as they are not affiliated with the academia, but with the industry instead. For them, public cloud infrastructures represent a natural source of computing power. On the other hand, there are examples of models (e.g. 0-D models), which do not require large-scale HPC for any individual simulation, but still need to run 10,000s of tasks, each ranging from seconds to minutes. Running them on a single machine would take hours or days, while simple parallelization could bring this time down to minutes, enabling faster iteration of the modeling process and a more interactive work mode. Such workloads are a good fit for serverless infrastructures, such as AWS Lambda or Google Cloud Run, which can run many short tasks in a highly elastic way, i.e. acquiring thousands of resources very quickly. 

In this paper we hypothesize that using a serverless computing model can be a practical and efficient approach to selected cases of VVUQ calculations. We show this on the example of the EasyVVUQ library, a popular tool for running VVUQ calculations, which we extend by providing support for AWS Lambda and Google Cloud Run serverless infrastructures. The resulting library, which we call CloudVVUQ, is then subject to evaluation, for which we use two real-world applications from the computational medicine domain, showing how they can be ported to the serverless infrastructure. Our experiments demonstrate the scalability of the proposed approach, by obtaining parallelism with 1500 tasks running concurrently, and achieving a 723-fold improvement in performance. 

The main contributions of the paper are as follows:

\begin{itemize}
    \item We describe the problem of VVUQ in computational medicine, its computational requirements and serverless infrastructure applicability;
    \item We create a prototype library (named CloudVVUQ) that accelerates sensitivity analysis of computational models, by using cloud serverless services such as the aforementioned AWS or Google Cloud solutions for parallel sample computation;
    \item We test the performance and behaviour of the serverless approach to sensitivity analysis in two real-world scenarios from computational medicine.
\end{itemize}

\section{Background}

\subsection{Serverless model}
In recent years the serverless model has become more efficient and popular. Serverless computing is a cloud execution model in which the cloud provider is responsible for provisioning and managing servers. One of the most important features of this model is auto-scaling, which means that higher demand triggers an increase in the amount of provisioned resources. A characteristic feature of serverless solutions is the pay-per-use billing model which often reduces costs compared to renting a server, and represents the optimal approach to matching resource allocation to problem size. However, it is not a perfect model for long-running jobs.

\subsection{Serverless computing services}
Cloud services that offer serverless computing have much in common with one another. Those services are built in an event-driven architecture and can respond to HTTP requests. Under the hood, there is often a Linux container that runs the application. In general, they are created to respond quickly, but sometimes before the first requests are processed the containers take a few seconds to initialize, which is known as a ``cold start''. The most common features are auto-scaling, high availability, concurrency limits, and security. We selected three services from the leading cloud providers which we regard as most suitable for the task. Each has its own strengths and weaknesses. The chosen services are listed below.
\begin{itemize}
    \item AWS Lambda \cite{AWS_Lambda}
    \item Google Cloud Functions (GCF) \cite{GCF}
    \item Google Cloud Run (GCR) \cite{GCR}
\end{itemize}
A more in-depth comparison is presented in Table ~\ref{table:cloud_services_comparision}.

Maximum execution times specified in the table are essential for this topic because they restrict the range of models that are compatible with serverless services. Container support is also important to enable model execution in customised environments and dependencies.
Languages such as Octave or Matlab are valued in the scientific and modelling communities, but they are not supported by default at the time of writing. AWS Fargate is a service that may enable computing models without a time limit in a serverless execution environment. It is a promising area for future work, but we elected to focus on the services listed above.

\begin{table*}[t]
\caption{Comparison of the selected cloud services.}
    \centering
    \begin{tabular}{|c|c|c|c|}
    \hline
        Service & AWS Lambda & Google Cloud Functions & Google Cloud Run\\ \hline
        Execution environment & Amazon Linux / Custom & Linux & Custom \\ \hline
        Supported languages & Java, Python, Node.js, Go, Ruby, C\# &  Node.js, Python, Go, Java, C\#, Ruby & Depends on execution environment \\ \hline
        Memory allocation & 128 MB - 10,240 MB & 128 MB - 8192 MB & 128 MiB - 32 GiB \\ \hline
        CPU allocation & Proportional to allocated memory & Proportional to allocated memory & 1 - 4 CPUs \\ \hline
        Disk space & Up to 512 MB & Uses memory & Uses memory \\ \hline
        Maximum execution time & 15 min & 9 min & 60 min \\ \hline
        Maximum instances & 1000 (default quota) &  Not specified / Unlimited & Over 1000 (region-specific) \\ \hline
        Deployment unit & Zipped code / Container image & Zipped code & Container image \\ \hline 
    \end{tabular}
    \label{table:cloud_services_comparision}
\end{table*}

\subsection{Computational Models}
Real-life systems are complex, and their behavior is difficult to predict.  Physically representing the relationships between the components is often impossible or cumbersome. 
Representing the problem mathematically, in an idealised form, reduces the crucial aspects to a set of formal numerical expressions known as a model. 
Mathematical models describe the system’s behavior and evolution using algebraic equations/inequalities, automata and state transition diagrams, probability distributions or systems of ordinary and partial differential equations, as explained in detail elsewhere \cite{Bungartz2014}. In the case of computational modelling, computers aid the modelling process. Typically, a model takes known information as input, while data produced by executing the model (referred to as a simulation) is the output. When the model is solved analytically, exact results are obtained using well-known analytical functions. 
On ther other hand, when the model is solved numerically, numerous methods are available. Those most relevant in our context include Euler and Runge-Kutta methods, and, for spatially distributed models, a finite difference/element approach \cite{Tekkaya2014}. Despite being subject to approximation, numerical solutions are typically attainable -- in contrast, many models do not have a known analytical solution.

To tackle any problem of interest, models should be as simple as possible, yet sufficiently complex to describe the real phenomenon being modelled. Care must be taken to define what is anticipated from the model at the start of the process. Verification, validation, and uncertainty quantification (VVUQ) are carried out to make sure the modelling choice is “right” for a given application. Verification is a process of investigating whether the model has been implemented correctly, both in terms of logic and coding -- in other words, whether the computer simulation model is consistent with the mathematical/conceptual model it was based on \cite{Murray-Smith2015}. Validation is a process which aims to determine if the model accurately represents the behaviour of the modelled system, i.e. if the model output matches observed data.
Another crucial and related component of model development is model sensitivity evaluation and quantification of uncertainty propagation from model inputs to model outputs.
Sensitivity analysis assesses the relative importance of different input factors and their contribution to the output uncertainty \cite{Saltelli2004book}, as described in the following section.

\subsection{Sensitivity analysis}
``Sensitivity analysis (SA) is the study of how the uncertainty in the output of a model (numerical or otherwise) can be apportioned to different sources of uncertainty in the model input''~\cite{Saltelli}. SA assesses the relative importance of model parameters. Ranking the parameters according to their impact on the output helps identify input parameters with minimal significance, which might then be set to a constant value, or \emph{screened}, reducing the overall dimension of the input parameter space. The computational cost of conducting the simulation is reduced, together with the complexity of the model.
This is relevant for many applications, for instance in medicine, where model outputs identify disease bio-markers \cite{Benemerito2022}.

There are two main types of sensitivity analysis - local and global. Local sensitivity analysis (LSA) evaluates the effect of a given input parameter on a given output parameter. Customarily, it relies on computing system derivatives of the form $S_j = \tfrac{\partial Y}{\partial X_j}$, where $X_j$ is the jth input parameter, and $Y$ is the output of interest \cite{Saltelli2004book}. Then, $S_j$ is the sensitivity index. In LSA, parameters are typically varied, one at a time, about a single reference state. Consequently, this does not provide information about parameter interactions. 

In global sensitivity analysis (GSA), all input parameters vary simultaneously. A common type of implementation is variance-based as described by Saltelli \cite{Saltelli2010}.
In such a case, the output's variance characterises the uncertainty of the output, caused by the input variation. For a GSA it is necessary to consider the appropriate range for all parameters, which is not a requirement in a LSA as the variation about the reference state is typically small. Consequently, GSA provides a more rigorous quantitative tool than LSA for inspecting the model.

In the area of computational models, SA consists mainly of a few steps. The first is the generation of different model inputs (samples), from the starting input data or parameter distributions. Examples of sampling methods include Stochastic Collocation, Monte Carlo, and Polynomial Chaos Expansion. Each one provides an $n$-dimensional input vector (where $n$ is the number of input parameters) for each simulation. After executing the model for all samples, the last step is to analyse the sensitivity. For example, one can plot chosen sensitivity indices \cite{Saltelli2010} on a treemap or a heatmap, to allow interpretation. Our implementation of such a workflow can be seen in Fig.~\ref{fig:client_side_workflow}.

Despite the apparently straightforward nature of the SA process, as outlined above, each execution of the model is usually computationally expensive and the dimension of its input parameter space tends to be very large. GSA, in particular, requires the generation of a statistically significant number of output samples to support reliable sensitivity metrics, such as Sobol indices \cite{Saltelli2010}; as a consequence, SA of even a moderately complex model is frequently prohibitively expensive.  

\subsection{Motivation}
Many fields have adopted serverless computation, but most perform scientific calculations using High-Performance Computing (HPC) infrastructures. Large, long jobs are not suitable for short-lived serverless containers, but some computational models do not require such resources and fit within the constraints of the serverless model.
Examples of such models can be found in ~\cite{Otta2022,UISS-TB}. The former requires low computational resources - a single computation takes less than a second on one CPU core. In the latter it is reported that the VVUQ process for the UISS-TB model requires over 600 simulations. Each simulation takes one minute using a modern CPU and 32GB RAM. 
In addition, the serverless approach speeds up the performance-heavy steps of the sensitivity analysis with parallel computation. As many companies and scientists lack access to HPC infrastructure, cloud computing provides an alternative option for users who do not own a sufficiently powerful computer. This paper explores how the serverless approach can be applied to perform VVUQ analysis on models relevant to the development of a digital twin. 

\section{Related Work}
\subsection{Generic libraries for serverless data processing}
The groundwork for serverless computational processing has been laid with the PyWren library \cite{PyWren}. This prototype Python library showed that it is possible to create a data processing system that inherits the elasticity and simplicity of the serverless model, using stateless functions with remote storage. PyWren users may not achieve the best parallel performance, but are significantly at an advantage compared to standalone workstations, while also reducing development time. PyWren uses AWS Lambda as the execution platform for serialized Python functions with the CloudPickle library. There is also an extension called NumPyWren that focuses on linear algebra. \cite{NumPyWren}. 
Recent developments in the scope of FaaS (Functions as a Service) applications for MapReduce-like computations are discussed in \cite{Lithops}. Lithops - an open-source Python library presented in this publication - can be described as a PyWren successor. It improves performance at both the invocation phase and the result collection phase, and adds many new features such as nested composability of functions or Jupyter Notebook integration. Lithops is a cloud-agnostic library which is probably its most important feature. It supports many cloud providers, and also offers support for Kubernetes \cite{Kubernetes}. 

\subsection{Serverless approach in science}
In \cite{ServerlessWorkflows} the authors discuss the advantages and limitations of serverless containers. Their work focuses on the usefulness of such solutions for scientific workflows. They present the results of performance and cost evaluation of chosen services in the CaaS (Container-as-a-Service) model (e.g. Cloud Run) and AWS Lambda. They conclude that using containers for short tasks (taking less than several minutes) is not recommended due to initial overhead. Furthermore, they agree that AWS Lambda is more cost-efficient than CaaS services, as long as the task matches the Lambda limits. Another insight from this publication is that a hybrid approach is feasible and effective (note that by ``hybrid'' the authors mean using AWS Lambda for fine-grained tasks while relying e.g. on AWS Fargate for more demanding tasks). The authors of \cite{funcx} developed a FaaS platform for scientific computing - funcX. Their work shows that it is possible to make use of diverse computing resources such as HPC and cloud, as well as to create a federated FaaS system. For testing purposes they used scientific case studies such as machine learning interference, real-time data analysis for High Energy Physics, Synchrotron Serial Crystallography and X-ray Photon Correlation Spectroscopy. There are other examples of using the serverless model for scientific or engineering applications, but VVUQ has not heretofore been supported in this way.

\subsection{VVUQ libraries}
Many frameworks and libraries have been created to facilitate VVUQ processes and provide necessary methods. Examples include SALib \cite{SALib}, Uncertainpy \cite{Uncertainpy}, UQTk \cite{UQTk} for Python and UQLab \cite{UQLab} for Matlab. We focus on EasyVVUQ, which is an open-source Python library designed to facilitate verification, validation and uncertainty quantification for a wide variety of simulations. The goal of EasyVVUQ is to make it easier to implement advanced VVUQ techniques for existing application codes or workflows \cite{EasyVVUQ}. It exposes these features mainly for simulation codes running on HPC. However, this library, like others mentioned earlier, does not support serverless execution by design. 

\section{CloudVVUQ library}
\subsection{Base library and design goals}
Our new CloudVVUQ library is based on EasyVVUQ because it provides the necessary sampling and analysis methods for VVUQ. A Campaign is the main object in the EasyVVUQ library, responsible for running simulations, results collection and storage, and retrieving data for analysis. This object has a method to insert external runs which we use to substitute other types of computation (HPC) by inserting results from cloud executions. In this way, we can compute samples in a serverless manner and use existing sampling mechanisms and analysis methods available for the Campaign. Although CloudVVUQ uses EasyVVUQ methods internally, it is still compatible with other Python libraries designed for sensitivity analysis (e.g. SALib) due to its use of inheritance functionality. There is no requirement to use EasyVVUQ code if the only goal is to distribute the sampling calculation on the cloud and to gather results. With this approach, a scientist does not need to translate existing code to use CloudVVUQ, but this option remains available for sample generation and post-simulation analysis. Our library is released on an open-source basis in Sano's Github organization~\cite{Sano_github}. 

The main design goals of CloudVVUQ are:
\begin{itemize}
    \item extending EasyVVUQ library with cloud computing
    \item support for multiple serverless infrastructures 
    \item achieving high parallelism and fault tolerance
    \item compatibility with other existing SA libraries
\end{itemize}

CloudVVUQ was primarily designed to extend EasyVVUQ by allowing sample computation on serverless cloud services. However, this does not limit its usage for VVUQ applications: its base Executor class is generic enough to support sending any json-type payload to cloud services. We also adapted three EasyVVUQ tutorial examples and received identical results using CloudVVUQ. 

\subsection{Tools used in the implementation}
CloudVVUQ distributes work to the cloud by sending HTTP requests asynchronously. To achieve this, we used the popular combination of two libraries, Aiohttp and Asyncio. The former provides all the necessary features to send HTTP/1.1 requests in a fast and secure manner with a great deal of customizability, while the latter is part of the Python Standard Library to handle asynchronous processing. To improve fault tolerance and deal with the unreliable nature of communication or processing errors, the Backoff library is used to add retry functionality. To handle authorization with AWS and Google Cloud, the respective SDKs (Software Development Kits) are integrated to sign requests using user-provided credentials. 

\subsection{Constraints on computational models and assumptions}
Not every model is suitable for the considered cloud services. The processing time of the most time-consuming sample must be within the specified timeout constraint. In addition, services such as Google Cloud Functions have a limited range of runtime languages. If the model language is not supported, then it is necessary to create a container image and deploy it to services such as Cloud Run or AWS Lambda. The deployed model must set optimal values for provisioned memory, number of vCPUs, and maximum instances limits. Otherwise, the cloud provider may throttle the calculations if the limit is reached, or a timeout will occur if the provisioned resources are insufficient.

In most cases, the user will be a scientist with little to no experience with cloud technology. Therefore, we assume that by following guidelines provided in the documentation, a scientist will be able to prepare the model for the deployment. Then a cloud administrator/DevOps would deploy the model to the service and provide the configuration details (url, credentials) to the scientist.

\subsection{Model deployment considerations}
\label{sec:deployment}
Deployment details differ between serverless services, but there are top-level similarities.
In order to successfully deploy a computational model to the cloud a scientist (or devops engineer) must first choose the suitable service. The next step is to prepare a service-compatible API in front of the model that can handle incoming requests, pass the sample data to the model, receive the output and return the response back to the client. Then, depending on the service deployment type, we pack the code in the zip file or create a container image. In the latter case we also have to create a Dockerfile, build and image and upload it to a container registry service. Finally, we follow the steps in the Cloud Console to create a function, specify its parameters and choose the container image or upload our zip package. 
There are many ways to deploy code to cloud services, e.g. using CI/CD pipelines, CLI, SDKs. The described steps present one of the possible implementations.

\subsection{CloudVVUQ architecture and simplified usage example}
\begin{figure}[b]
    \centering
    \includegraphics[scale=0.73]{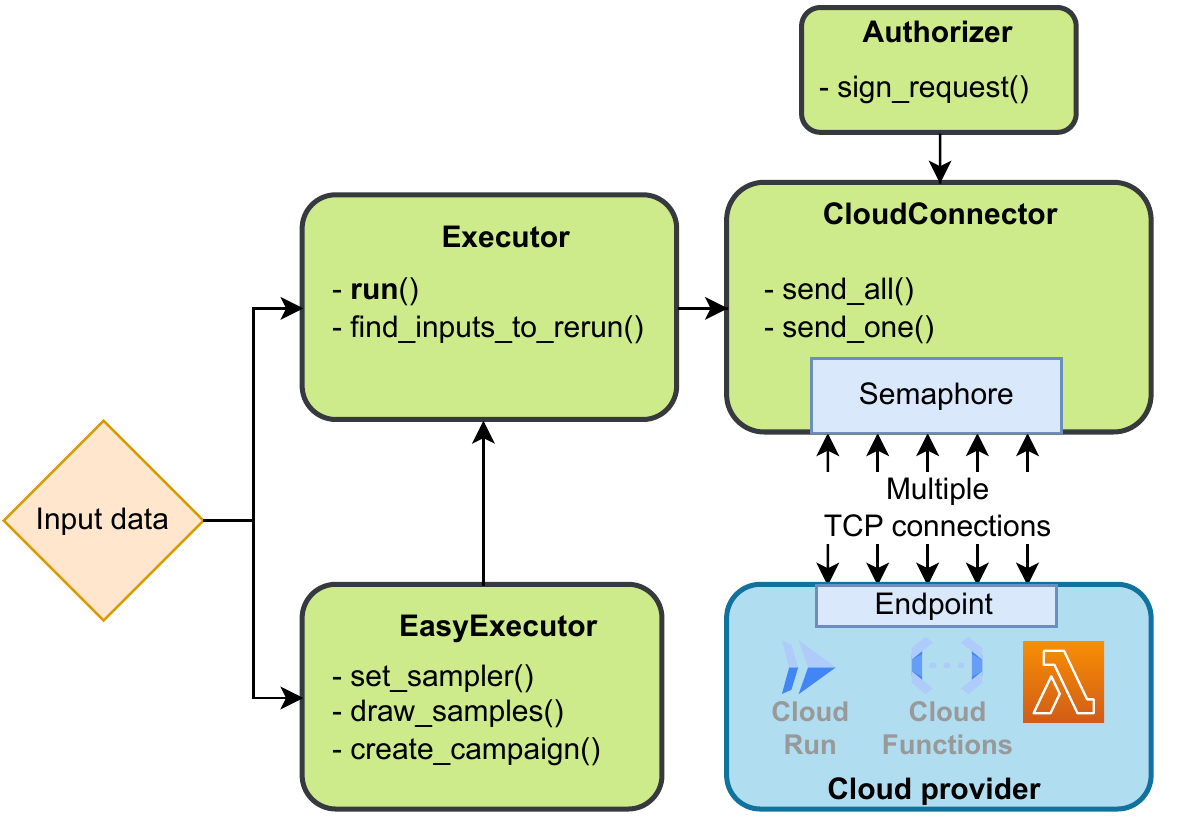}
    \caption{Diagram of CloudVVUQ architecture and classes.}
    \label{fig:cloudvvuq_architecture}
\end{figure}

The CloudVVUQ architecture is presented in Fig.~\ref{fig:cloudvvuq_architecture} and the API details can be found in the documentation in the repository.

To use CloudVVUQ users do not need to change their existing EasyVVUQ code in any significant way - this was one of the design goals. The example below shows a simplified CloudVVUQ script which can be used as a template.
\begin{lstlisting}
import easyvvuq as uq
from cloudvvuq import EasyExecutor

load_credentials()

params = define_parameters_and_ranges()
vary = define_variance_distributions()

sampler = uq.sampling.SCSampler(vary)
executor = EasyExecutor(url)
executor.set_sampler(sampler, params)
samples = executor.draw_samples()

executor.run(samples, max_load=256)

campaign = executor.create_campaign()
campaign.apply_analysis()

results = campaign.get_last_analysis()
results.plot_sobols_treemap()
\end{lstlisting}

\section{CloudVVUQ workflow}
\subsection{Client-side workflow description}
\begin{figure}[t]
    \centering
    \includegraphics[scale=0.51]{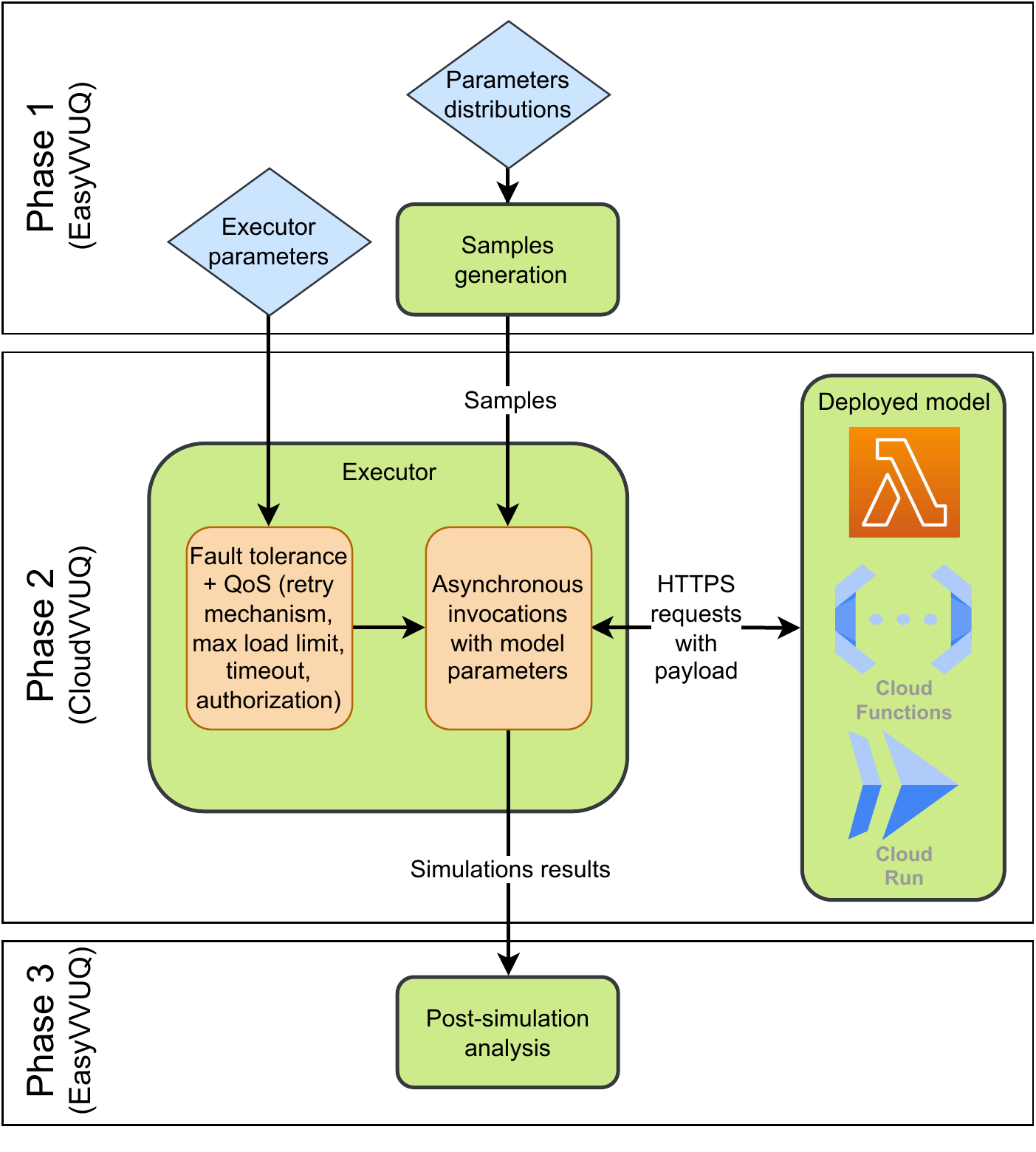}
    \caption{Diagram of client-side workflow}
    \label{fig:client_side_workflow}
\end{figure}

Once the model has been deployed to the cloud service according to the requirements mentioned above, it is time to send samples for computation using the CloudVVUQ library. The CloudVVUQ workflow, presented in Fig.~\ref{fig:client_side_workflow}, consists of three phases. The first is to generate samples and prepare them for transport. If we already have samples from some external source, we can use them instead of generating new ones. Then, in the following phase, we use asynchronous processing to keep sending multiple requests to an endpoint with a deployed model in the cloud. This step is inspired by the MapReduce programming model. The detailed algorithm used in the library is presented as Algorithm \ref{alg:client_side_algorithm}. The most important aspect of the algorithm is that the client must use fault tolerance mechanisms, such as retries, to handle errors and connection issues and limit the number of outgoing requests to an endpoint. The final phase is to perform an analysis based on simulation data. Our library provides the user with EasyVVUQ methods created for this task.

\begin{algorithm}[tb]
    \caption{CloudVVUQ client-side workflow}
    \label{alg:client_side_algorithm}
    \begin{algorithmic}
    
    \State $credentials \gets $ Load user credentials 
    \State $samples \gets $ Generate samples from distributions and using a given sampling method
    \State $requestsQueue \gets $ Prepare samples to send
    \State $semaphore \gets $ counting-semaphore to control the number of requests sent concurrently to the cloud
    \State $begin$ Async Thread (length($requestsQueue$))
    
    \While{$requestsQueue$ is not Empty} 
        \If{There are no responses to process} 
            \If{$semaphore$ is not full} 
                \State $semaphore = semaphore + 1 $ 
                \State $request \gets $ Pop request from queue 
                \State $request \gets $ Sign $request$ using $credentials$
                \State Send $request$ to given endpoint
                \State $Release$ Async Thread
            \ElsIf{$semaphore$ is full} 
                \State \textbf{continue} 
            \EndIf
        \ElsIf{There are responses to process} 
            \State $response \gets $ Load request response 
            \If{$response$ status != 200 $or$ Error code} 
                \State Append $request$ again to $requestQueue$
            \ElsIf{$response$ status == 200} 
                \State $result \gets $ $response$ payload 
                \State Save $result$ to file
                \State $end$ Async Thread 
            \EndIf
            \State $semaphore = semaphore - 1 $ 
        \EndIf    
    \EndWhile

    \State $simulationsResults \gets$ Load results from files
    \State Analyze $simulationsResults$

    \end{algorithmic}
\end{algorithm}

\subsection{Cloud provider-side workflow description}

\begin{figure}[tb]
    \centering
    \includegraphics[scale=0.55]{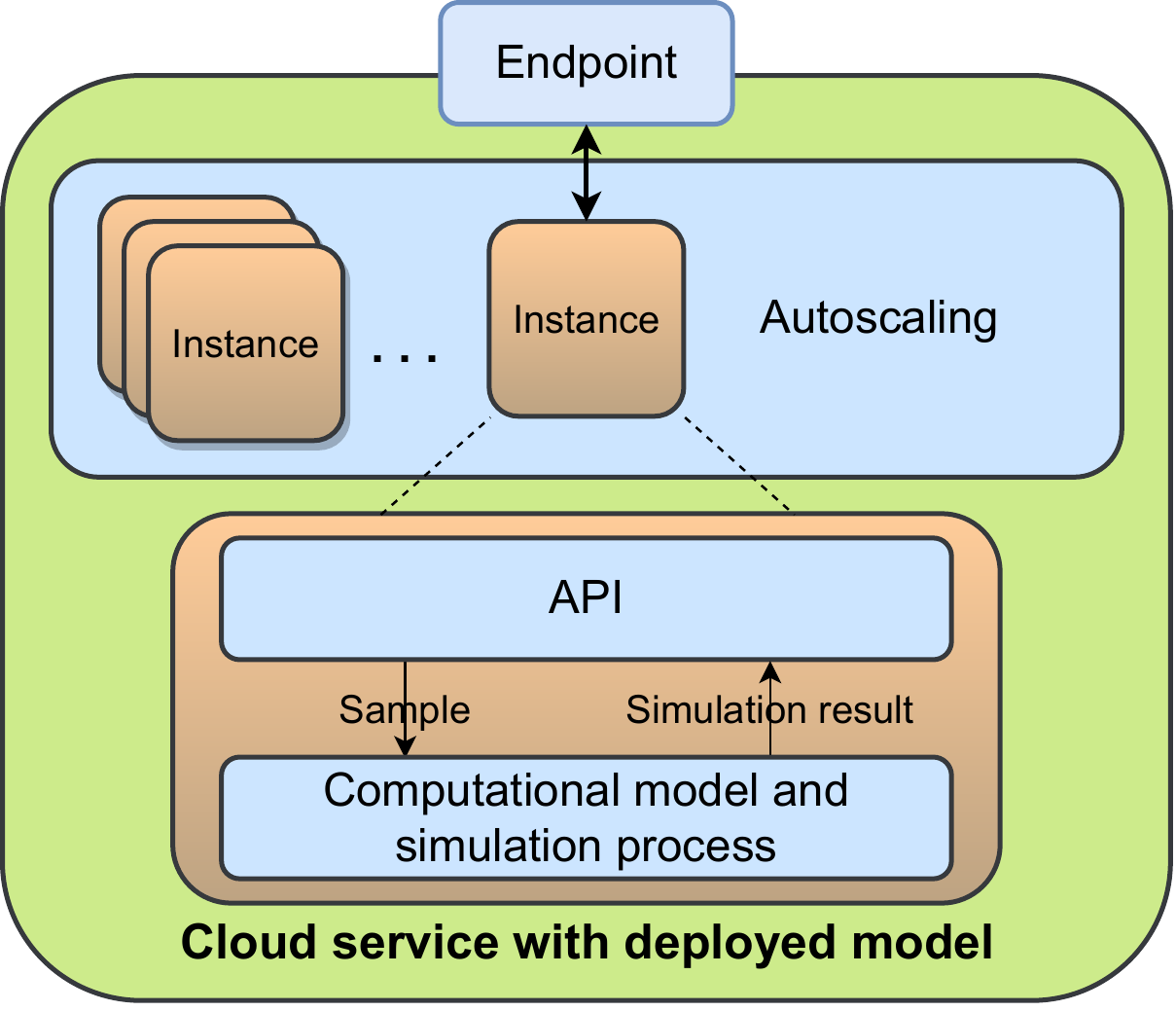}
    \caption{Diagram of cloud provider-side workflow}
    \label{fig:provider_workflow}
\end{figure}

The CloudVVUQ library requires the computational model to be deployed in the cloud. The model must be adapted for this purpose (see section \ref{sec:deployment}). Each cloud provider requires that the deployed function or container have an API for receiving requests and returning responses. Usually, for every request received at a given endpoint, the cloud provider starts a new container or reuses containers that are not processing anything at the moment. Therefore, computational models need to expose a method to calculate only one given sample that will be sent with the request. The pseudocode for this simple workflow is presented in Algorithm \ref{alg:provider_side_algorithm} and Fig.\ref{fig:provider_workflow}.

\begin{algorithm}[tb]
\caption{CloudVVUQ cloud provider-side workflow}
\label{alg:provider_side_algorithm}
\begin{algorithmic}
    \State $request \gets $ Receive request 
    \State $sample \gets $ Extract parameters from $request$ payload
    \State $process \gets $ Set up a new process with computational model simulation
    \State $results \gets $ Pass $sample$ to $process$ and wait for results
    \State $response \gets $ Create request with simulation $results$ as a payload
    \State Send $response$ to the client
\end{algorithmic}
\end{algorithm}

\subsection{Sample generation}
The EasyVVUQ library provides a useful API to generate samples. The user can specify the distribution and range for each parameter and pass it on to one of the many sampling methods implemented in this library. Depending on the sampling method and method-specific arguments, the number of samples may vary. The result is an array of samples ready to be calculated.
Other libraries, such as SALib, provide an alternative and are based on similar logic. 

\subsection{Concurrency and limits}
To achieve maximum speed-up of sample calculations, it is necessary to keep our cloud deployment saturated with samples. Cloud services usually have concurrency and maximum instances limits, and it is considered good practice to customize these limits during deployment. Containers, unlike 1st generation Cloud Functions, can process more than one request at a time. Therefore, it is necessary to set an optimal limit for concurrent processing to avoid overloading the container. Container concurrency and maximum instance limit together give us an upper bound for the maximum concurrency that we can achieve with CloudVVUQ. 

The client has other limits that we need to be aware of. One of them is the number of open TCP connections, which varies between individual computers and operating systems. CloudVVUQ uses the HTTP/1.1 protocol to send requests that require a TCP connection for each outstanding request, resulting in a connection pool size which runs into the hundreds or thousands. Another consideration is the speed of the client-side network, which can be a communication bottleneck. The final aspect relates to the CPU performance on the user's computer. Taking all these into account, we obtain another upper bound for maximum concurrency we can achieve - this time, from a client-side perspective.

\subsection{Fault tolerance}
Communication problems occur -- the server can drop the TCP connection, a packet may get lost, and the response status code may be other than 200 (success). This cannot cause stopping or terminating a workflow. To address this issue, a retry mechanism has been implemented. Whenever one of those situations arises, another request is en-queued to send with the same payload. Clearly, this should not happen indefinitely, therefore, we can specify a maximum number of tries. The downside of retries is that, in the worst-case scenario, they can significantly slow down the simulation results. An example of this could be a long-to-compute model with all samples being computed in parallel when a retry happens, then computation time is at least doubled (assuming each sample processing time is very similar). 

Client-side timeouts are also essential to handle situations in which the TCP connection is kept alive, but no request was returned in the specified time, e.g. maximum processing time for a given service. 

As mentioned above, cloud deployment has a limit for concurrent requests. The total number of samples we can calculate on the cloud is simply the number of instances multiplied by the instance concurrency limit, which for AWS Lambda or 1st Generation Google Cloud Functions is one request. Cloud Run container can process more than 1 request at a time, but it is only viable if the model is not resource-heavy and load-tuning tests verify this. In order not to overload the cloud deployment with requests, the client library uses a counting semaphore mechanism (see Algorithm~\ref{alg:client_side_algorithm}) to limit sent requests. In this way, we ensure that we process as many samples as possible without unnecessary errors from the cloud service and improve overall stability.

\section{Experiments}
The goals of the experiments are the following:
\begin{itemize}
    \item to test the CloudVVUQ with real-life applications from the computational medicine field,
    \item to show the scalability and parallel computing performance,
    \item to demonstrate fault tolerance mechanisms,
    \item to identify the performance limits and bottlenecks.
\end{itemize}

To test our approach we used two real-world computational models.
The first is a Lower Limb Haemodynamics model. Its calculation is relatively fast (150 ms). The second model is a modified CellML Ten Tusscher, Nobel, Nobel, Panfilov \cite{TNNP_repo} model of ventricular tissue with a much longer calculation time (4 min). By choosing these two models, we can test the CloudVVUQ and the serverless approach in two scenarios differing in granularity of computing tasks. 

\subsection{Testing platform and configuration}
For testing purposes the CloudVVUQ library was installed on a modern laptop computer with specification listed below. Also a high-speed, High-throughput connection was used to minimize possibility and impact of network bottleneck. 
\begin{itemize}
    \item CPU - Intel Core i7-10510U
    \item RAM - 16 GB
    \item OS - Windows 10 Pro
    \item Connection - 960 Mb/s download, 900 Mb/s upload
    \item TCP connections limit - ~2000
\end{itemize}

The full configuration of model deployments, along with testing parameters is specified in Table ~\ref{table:test_config}. We decided to include cold starts in the tests for more realistic results and links to the source codes for reproducibility purposes.

\begin{table}[h]
\caption{Description and configuration of tests.}
    \centering
    \begin{tabular}{|c|c|c|}
    \hline
        Model & \begin{tabular}[c]{@{}c@{}}Lower Limb\\Haemodynamics\end{tabular} & Ten Tusscher et al \\ \hline
        Written in & Python & Octave/Matlab \\ \hline
        Cloud Service & \begin{tabular}[c]{@{}c@{}}Cloud Functions\\(1st gen)\end{tabular} & Cloud Run \\ \hline
        Average simulation time & 150 ms & 4 min \\ \hline
        Memory allocation & 2 GB & 512MiB \\ \hline
        CPU allocation & 1 vCPU & 1 vCPU \\ \hline
        Instance concurrency & 1 & 1  \\ \hline
        Maximum instances & 4500 & 1200  \\ \hline
        Region & europe-west1 & europe-west1  \\ \hline
        Provider-side timeout & 30 s & 25 min \\ \hline
        Client-side max load & 600 & 500 \\ \hline
        Max TCP connections & 1200 & 1000 \\ \hline
        Test size & 54272 samples & 1000 samples\\ \hline 
        \begin{tabular}[c]{@{}c@{}}Payload size\\(send / receive)\end{tabular} & 2.6 KB / 7.2 KB &  7.25 KB / 1.8 KB \\ \hline
        Warm-up & No & No \\ \hline 
        Link to the source code & https:// ... & https:// ... \\ \hline
    \end{tabular}
    \label{table:test_config}
\end{table}

\subsection{Lower Limb Haemodynamics model description}

A steady-state 0D model of lower limb haemodynamics was taken from \cite{Otta2022}. It is a reduced-order model based on the electric-hydraulic analogy, where the vascular network is represented by an electrical circuit. The model characterises the global pressure-flow distribution in blood vessels of the lower limb. It is informed by 42 input parameters (vessel radii) and produces 50 output parameters (flow rates in vessels). The model was used to perform a formal global sensitivity analysis with input parameter deviation of up to 50\%. In the original paper, the sampling method used was Saltelli's extension of the Sobol sequence implemented in the SALib library. More than 100,000 samples were required to achieve convergence of the first-order Sobol indices. For testing purposes we used a smaller but still meaningful set of 54272 samples.

\subsection{Ten Tusscher model description}

The Ten Tusscher, Nobel, Nobel, Panfilov model \cite{TNNP} is an established model of the human heart ventricle cells designed to represent the electrical potential in the cell. It is based on the transfer of ions through the cell membrane. The code representing the model was taken from the CellML repository \cite{CellML_repo} \cite{Cellml_article} and modified to run the full heart cycle and output averages and/or extremes of values of some model states. The model was used to perform a local sensitivity analysis. Three (out of 63) main model parameters were sampled from a normal distribution resulting in up to 45\% deviation, the others were set to default values. The polynomial order parameter for Stochastic Collocation Sampler was set to 9, effectively giving us 1000 samples.  

\subsection{Testing Lower Limb Haemodynamics model}
\begin{figure}[H]
    \centering
    \includegraphics[scale=0.29]{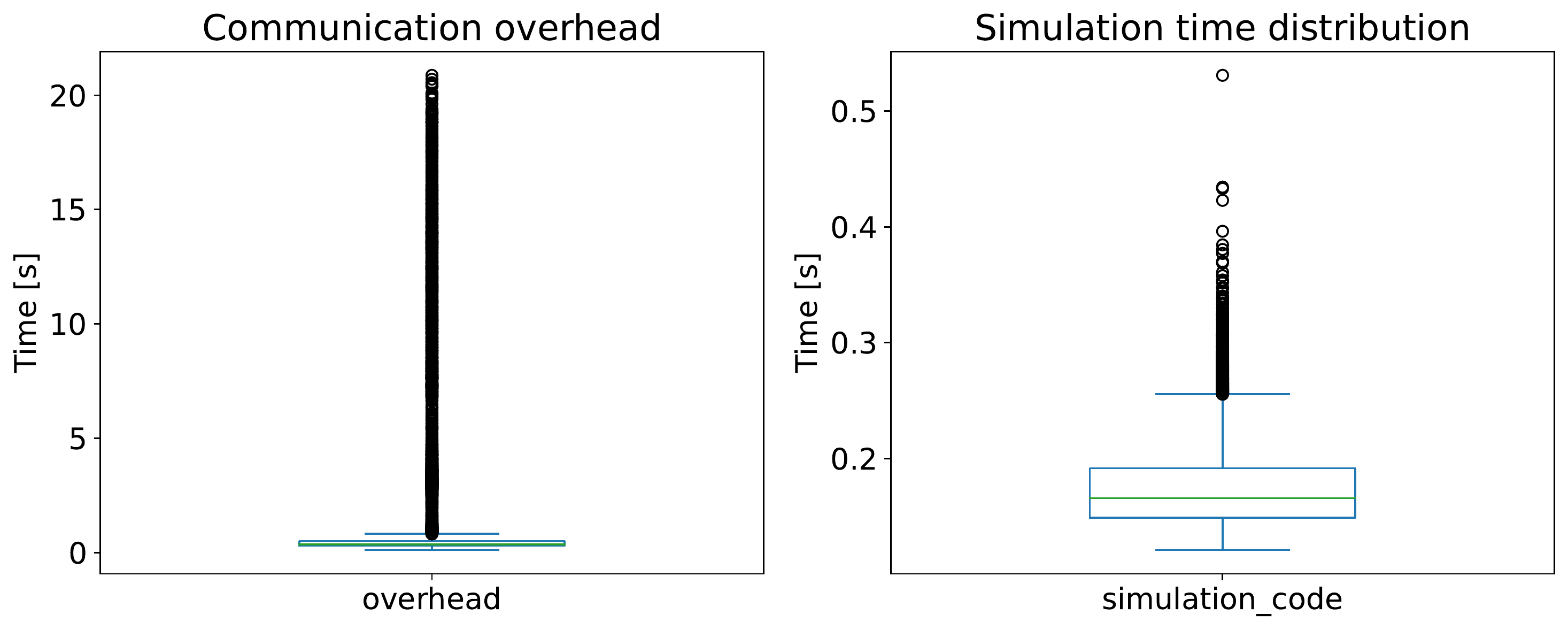}
    \caption{Communication overhead and simulation time for the Lower Limb Haemodynamics model executions using GCF.}
    \label{fig:llm_distributions}
\end{figure}

This model (implemented in Python) was used to test the performance of CloudVVUQ and Google Cloud Functions for relatively quick calculations. Computational models rarely have such short execution times but this case is interesting from the implementation perspective and by using the CloudVVUQ library the user can work in almost interactive mode -- thanks to reduction of time-to-result we can achieve by parallelism and dynamic provisioning capabilities of serverless infrastructure.   Fig.~\ref{fig:llm_distributions} shows distributions for communication overhead. About 96\% of the results have overheads of less than 1 second. About half of the remaining 4\% - seen as outliers in Fig.~\ref{fig:llm_distributions} - can be attributed to starting overhead, such as cold starts, which can add even 2 seconds overhead or setting up one TCP connection for each concurrent request as required by the HTTP/1.1 protocol. 

\begin{figure}[H]
    \centering
    \includegraphics[scale=0.29]{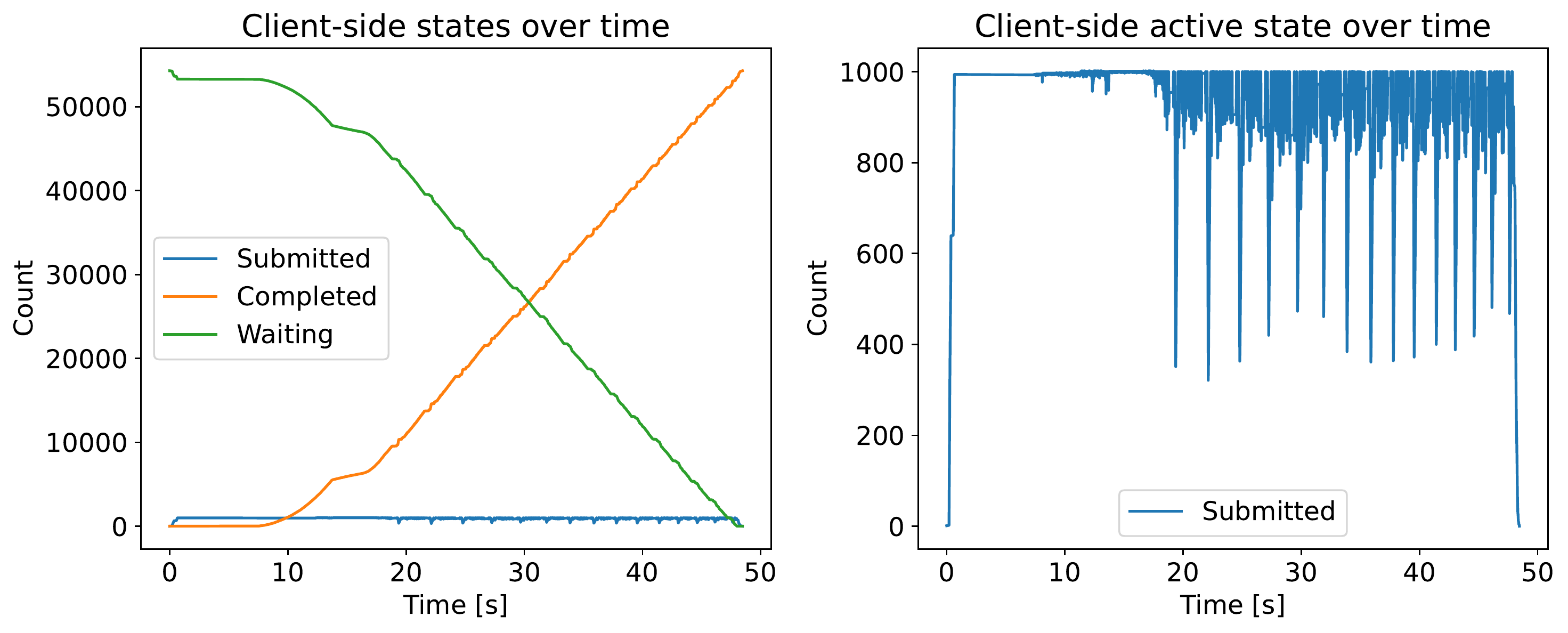}
    \caption{Client-side samples' states for the Lower Limb Haemodynamics model executions using GCF.}
    \label{fig:llm_client_side_states}
\end{figure}

Although the maximum client-side load limit was 1000 concurrent requests, we see in Fig.~\ref{fig:llm_client_side_states} that the testing platform was not able to fully utilize the service. The load kept oscillating between 400 and 1000 sent requests (submitted states). The explanation is that responses have precedence over sending new requests. An improvement for quick simulations would be to merge a few samples into one request, split them in the function, then calculate each sample sequentially or in parallel and finally return a response with respective outputs. 

\begin{figure}[H]
    \centering
    \includegraphics[scale=0.29]{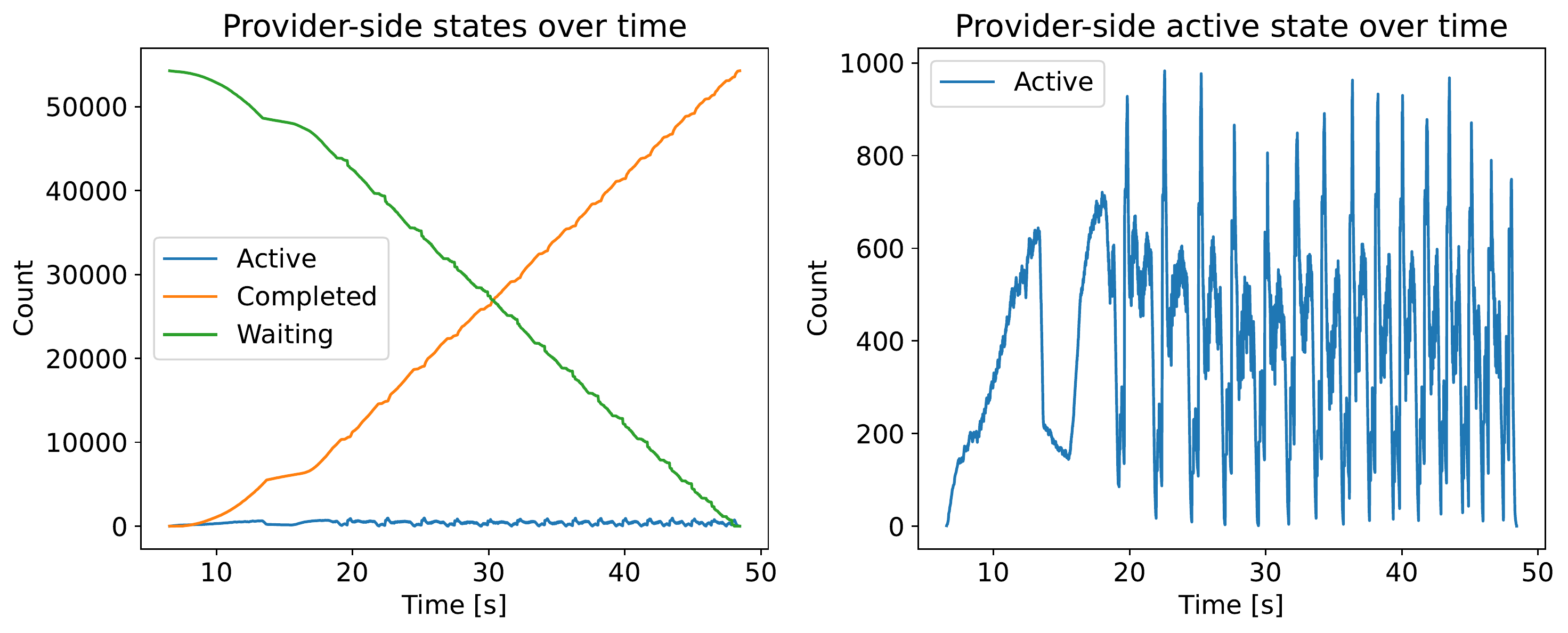}
    \caption{Cloud provider-side samples' states for the Lower Limb Haemodynamics model executions using GCF.}
    \label{fig:llm_provider_side_states}
\end{figure}

\begin{figure}[H]
    \centering
    \includegraphics[scale=0.29]{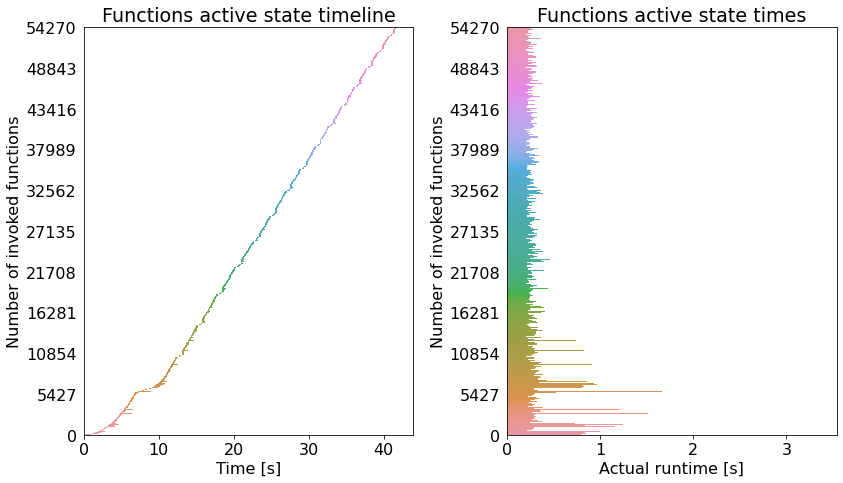}
    {\footnotesize Plotted results are sorted by the time a container received a request \par}
    \caption{Runtime plot for the Lower Limb Haemodynamics model executions using GCF.}
    \label{fig:llm_runtimes}
\end{figure}

The maximum concurrency achieved on the provider side was 983 simulations. Fig.~\ref{fig:llm_provider_side_states} also shows high variability in the number of samples processed simultaneously. This is due to the transport latency and the very fast simulation time. Despite this, in Fig.~\ref{fig:llm_provider_side_states} we can see a steady increase in completed simulations, and Fig.~\ref{fig:llm_runtimes} shows that no throttles occurred during the entire workflow.

Performing those calculations locally could easily take three hours (depending on the platform). The total serverless workflow execution took about 48 seconds and the sum of all simulation times (excluding overhead) in the cloud was about 2.6 hours. Using CloudVVUQ on our platform, we can speed-up the sample calculation phase at least 195 times for this quick-to-compute model. During the workflow a total of 7 retries occurred for requests with IDs = [1052, 882, 528, 787, 358, 306, 522] which were successfully executed last. Such low ID values suggest that the reason is related to initial provisioning of instances in the cloud. 

\subsection{Testing the Ten Tusscher model}

\begin{figure}[H]
    \centering
    \includegraphics[scale=0.29]{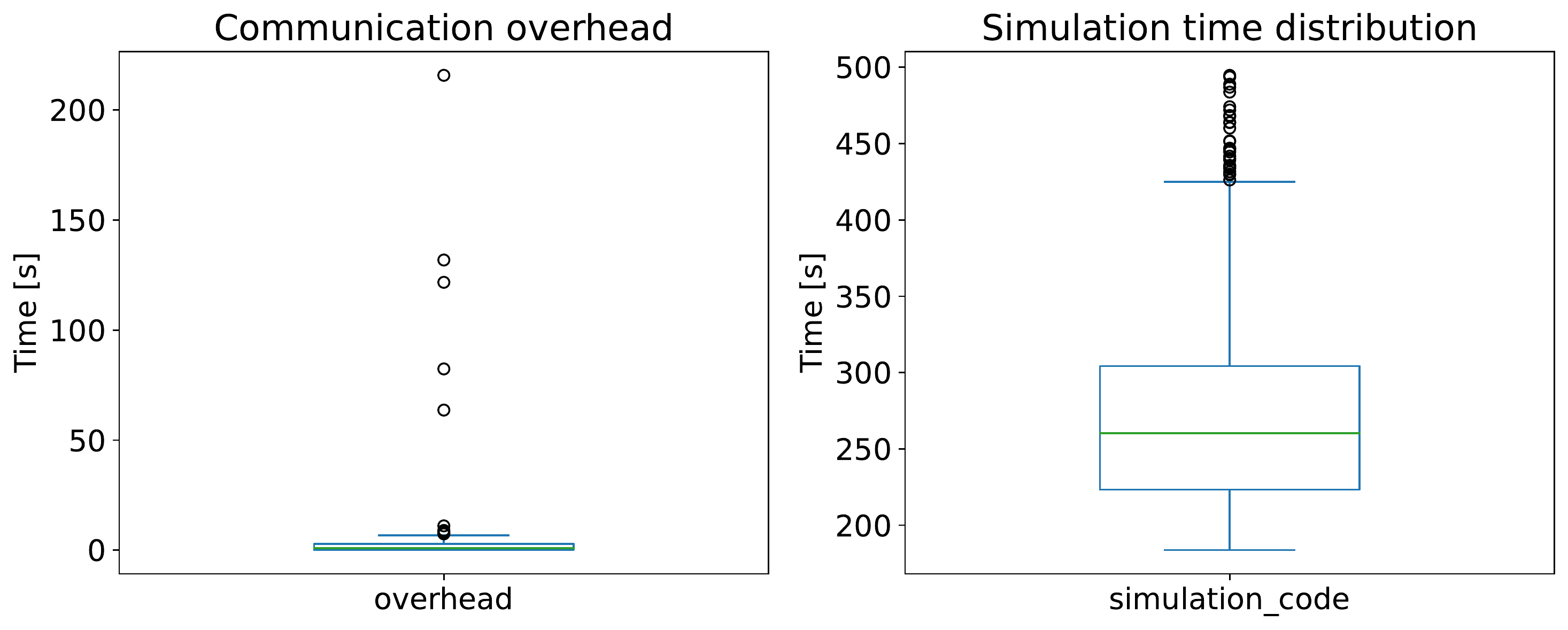}
    \caption{Communication overhead and simulation time for the Ten Tusscher model executions using Cloud Run.}
    \label{fig:cellml_distributions}
\end{figure}

This model is written in Octave, thus a container is required for the deployment, making Cloud Run a suitable service. The mean simulation time is about 4 minutes, therefore our testing platform should be able to saturate the cloud with requests and achieve an almost arbitrary number of parallel simulations (compared to the previous model). As we see in Fig.~\ref{fig:cellml_distributions}, the mean communication overhead is 2.2s and higher times can be linked to cold starts, with some outliers waiting for over 50s to be processed. 

\begin{figure}[H]
    \centering
    \includegraphics[scale=0.29]{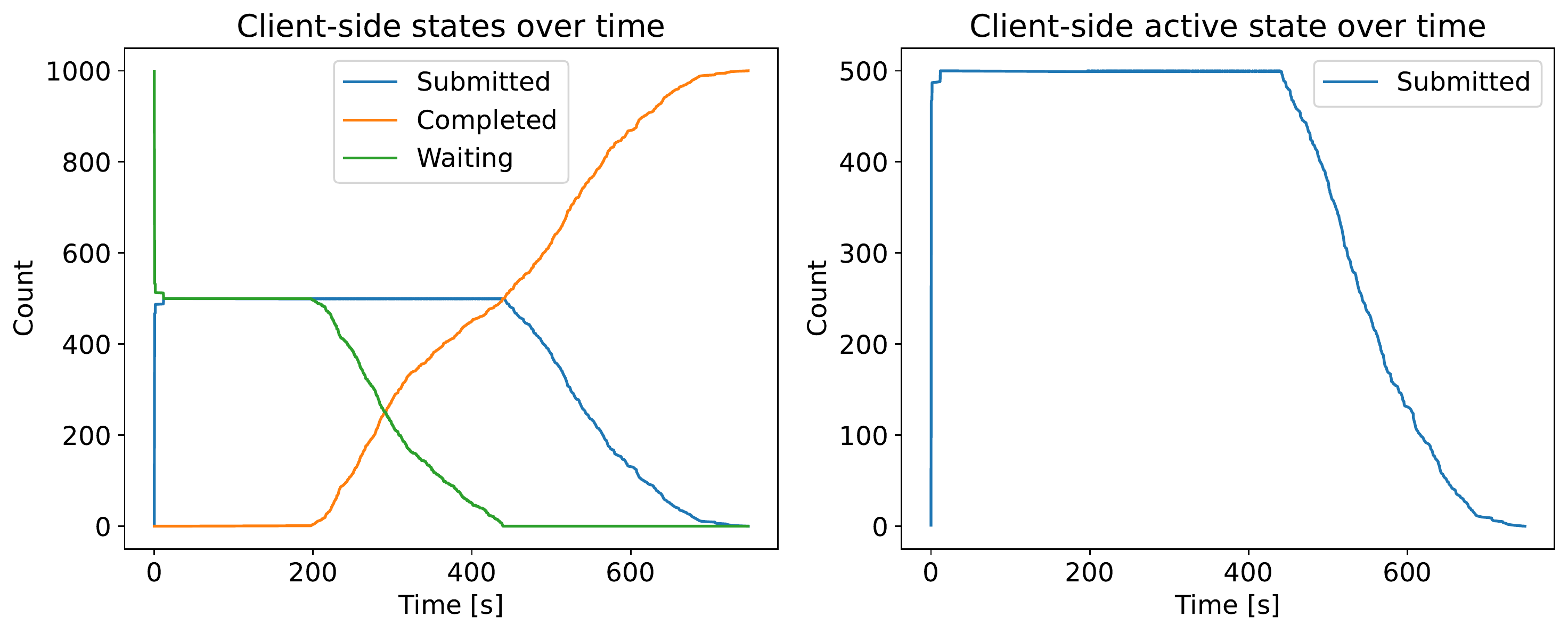}
    \caption{Client-side samples' states for the Ten Tusscher model executions using Cloud Run.}
    \label{fig:cellml_client_side_states}
\end{figure}

\begin{figure}[H]
    \centering
    \includegraphics[scale=0.29]{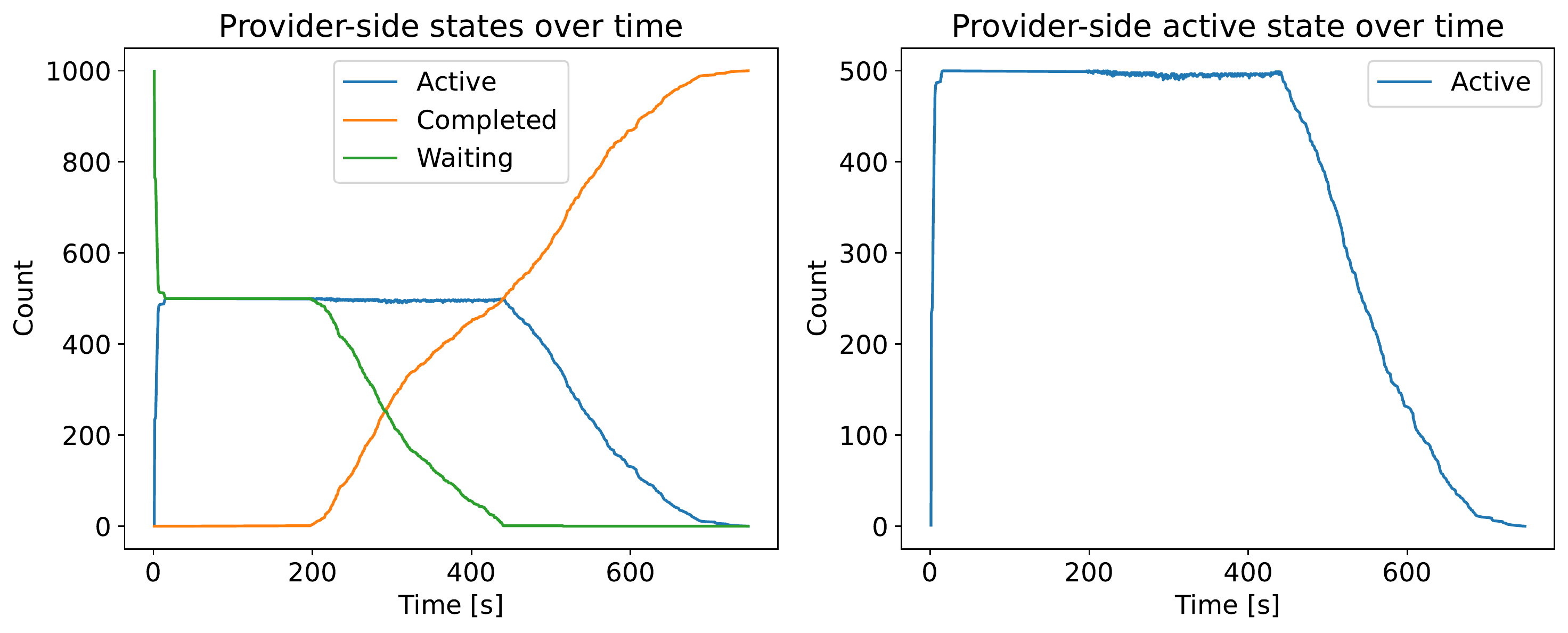}
    \caption{Cloud provider-side samples' states for the Ten Tusscher model executions using Cloud Run.}
    \label{fig:cellml_provider_side_states}
\end{figure}

\begin{figure}[H]
    \centering
    \includegraphics[scale=0.29]{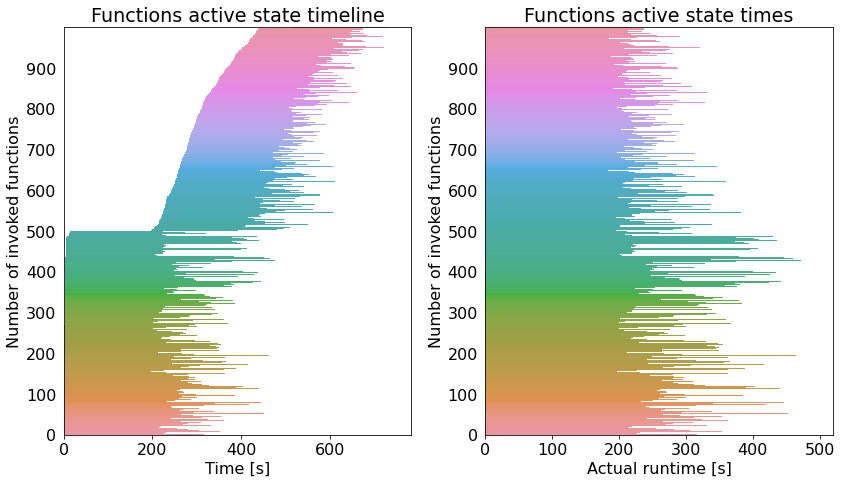}
    {\footnotesize Plotted results are sorted by the time a container received a request \par}
    \caption{Runtime plot for the Ten Tusscher model executions using Cloud Run.}
    \label{fig:cellml_runtimes}
\end{figure}

Fig.~\ref{fig:cellml_runtimes} shows that as soon as a response is received, another request is sent to make full use of the service without exceeding the limit. Because this model takes longer to compute and the concurrency limit is lower compared to the Lower Limb Model test, this platform has no problem with keeping the maximum load of 500 and the provider side reflects that in Fig.~\ref{fig:cellml_provider_side_states}. Also it shows that the service almost immediately provisioned the required number of containers. The similarity of Fig.~\ref{fig:cellml_provider_side_states} and Fig.~\ref{fig:cellml_client_side_states} means that the communication overhead is negligible in this case. Because this model takes longer to compute, we could achieve greater speed-up than the previous model. The calculation time was 366 times shorter than the sum of only simulation times in the cloud, which approximates sequential processing.

\subsection{Full parallelization experiment}

\begin{figure}[H]
    \centering
    \includegraphics[scale=0.29]{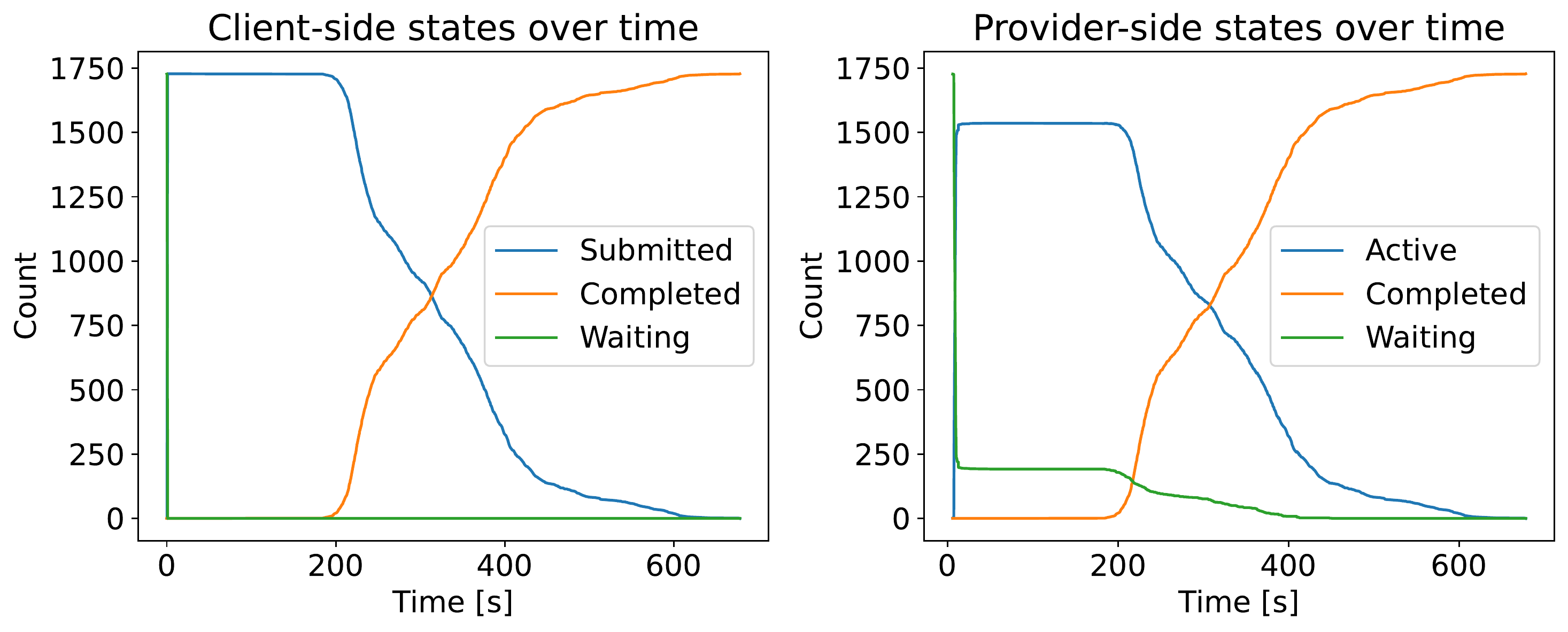}
    \caption{Samples' states for fully parallel execution test of Ten Tussher model using Cloud Run.}
    \label{fig:cellml_client_provider_side_states_1728.pdf}
\end{figure}

It is possible to calculate all samples in parallel if the current resources and limits allow it. We performed another test for the Ten Tusscher model with the same configuration except for an increased number of samples (1728) along with client- and provider-side limits above 2000. In this way we achieved up to 1536 active containers, as seen in Fig.~\ref{fig:cellml_client_provider_side_states_1728.pdf} and a 723-fold improvement in sample processing time.

\subsection{Testing Ten Tusscher model on AWS Lambda}
\begin{figure}[h]
    \centering
    \begin{subfigure}[h]{0.22\textwidth}
        \centering
        \includegraphics[width=1\textwidth]{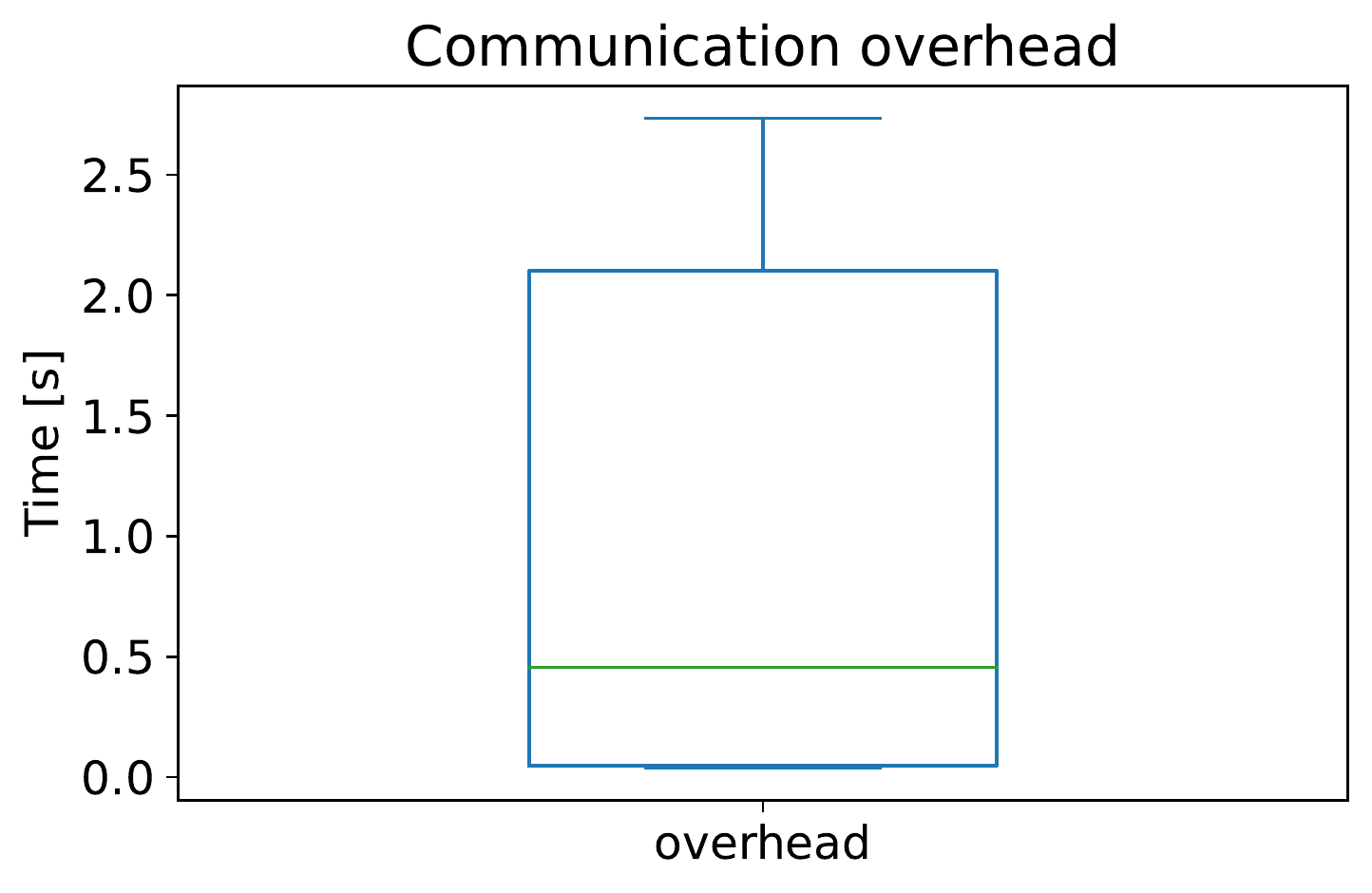}
        \caption{Overhead distributions}
        \includegraphics[width=1\textwidth]{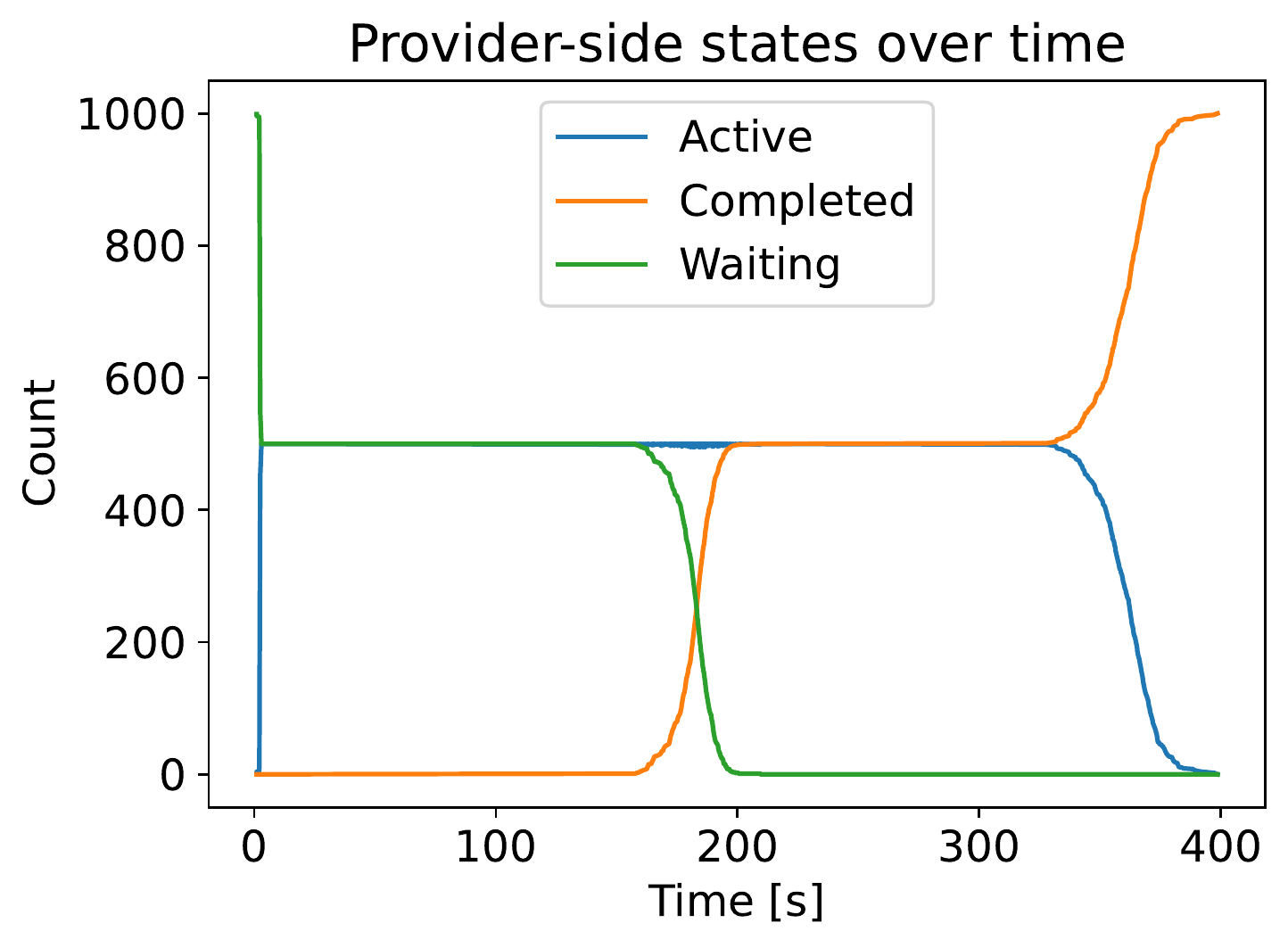}
        \caption{Provider-side samples' states}
    \end{subfigure}
    ~ 
    \begin{subfigure}[h]{0.22\textwidth}
        \centering
        \includegraphics[width=1\textwidth]{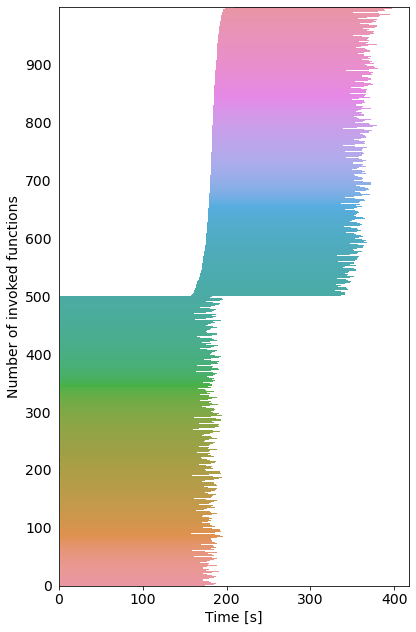}
        \caption{Runtime plot}
    \end{subfigure}
    \caption{Test results for Ten Tusscher model executions using AWS Lambda}
    \label{fig:cellml_tests_aws}
\end{figure}

The last test is performed once again on a Ten Tusscher model, this time using the AWS Lambda service. We recreated the configuration to be as similar to the one defined in Table ~\ref{table:test_config}. In particular, 1800MB of memory was provisioned. The model uses less than 300MB, however, memory in AWS Lambda is strictly tied to provisioned CPU and our value grants approximately 1 vCPU \cite{Lambda_1vCPU}. In Fig.~\ref{fig:cellml_tests_aws} we can see nearly perfect execution results. Almost instantly, the required number of instances was provisioned. Overhead remained stable and minimal - the first half of requests involved cold-start overhead of approximately 2 seconds, while the second half reused warm instances. No outliers occurred. The service was used at 100\% - up to the configured concurrency limits. 

\subsection{Summary of Results}
The test's outcome is a success and we have completed all our testing goals. We performed computation of samples using serverless services with a significant speedup. The full parallelization experiment shows that we can easily control the number of parallel computations and thus achieve results much faster. Moreover, we could identify the influence of network communication and short simulation time on performance. Most issues that may occur during the second phase involve the server closing the TCP connection and the client receiving HTTP-429 (too many requests) statuses, usually due to an initial lack of instances in the ready state. This was probably the case for the Lower Limb Model test where retries occurred and the retry mechanism proved to be a crucial component. 
Tests showed that AWS Lambda provides the best performance compared to other services. The authors of \cite{copik2021sebs} reached the same conclusion and also identified that Cloud Functions experience reliability issues.

There is a place for future optimization, e.g. fine-tuning containers for multiple requests processing, using Google Cloud Functions 2nd generation that supports concurrent request processing, improving client-side configuration (optimal number of TCP connections), finding the best retry strategy, and performing model-specific optimization. 

\section{Discussion}
Here we briefly summarize our general observations and lessons learned from our prototype and experiments.
\subsection{Serverless vs HPC}
We proposed serverless computing as an alternative to HPC for embarrassingly parallel tasks of VVUQ calculations. One of the main advantages of FaaS services is that the submitted work is processed almost instantly, while jobs submitted to the SLURM system are placed in a queue and wait for resources, which can take a long time if the cluster is under heavy load. A small disadvantage of our approach is that we cannot cancel submitted and ongoing computations to reduce costs and usage, while HPC supports job cancelling. Nonetheless, in many cases is it much easier to obtain access to cloud computing in comparison to securing a grant on a supercomputer. 

\subsection{HTTP protocol issues}
HTTP/1 is a well-established and mature protocol but has an important drawback called 'head-of-line blocking', which means that only one request can occupy the connection. HTTP-pipelining introduced in HTTP/1.1 attempts to mitigate the issue. Launching 500 TCP connections to achieve 500 concurrent simulations is viable if there is no alternative, but it is very excessive and suboptimal. Our initial tests with the HTTP/2 protocol show promising results - with a single TCP connection we can compute Lower Limb Haemodynamics model's samples only 5.5 times slower than using HTTP/1.1, and this can be further improved by distributing samples among the optimal number of new processes. Furthermore, with HTTP/2 there are much fewer error responses (HTTP 500 or 429 codes) than when using HTTP/1.1. Unfortunately, we cannot expect that every endpoint will support this protocol (e.g. AWS Lambda requires API Gateway for this purpose).

\subsection{Other observations on serverless platforms}
During the testing phase we encountered some interesting behavioral patterns of cloud platforms. 

The first one was that after a period of inactivity (e.g. 1 month) within the deployment we should expect degraded initial performance regarding scalability and a high number of errors, due to the lack of provisioned instances. In AWS this is explained by the fact that configured external resources are reclaimed and the function must be recreated. GCP deployments also experience this problem, but soon after restoring traffic (within a few dozen seconds) everything again works as intended. 

The second issue we observed during experiments is that the Cloud Run service quite often under-provisioned containers for highly parallel (over 700 instances) simulation workflows. Instead of provisioning 1 container for each request as expected, the active state plots show that the service provisioned approximately 90\% of the required value (submitted), as seen in Fig.~\ref{fig:cellml_client_provider_side_states_1728.pdf}.

We estimate that running our VVUQ experiments using serverless services was 2-2.5 times more expensive than using a virtual machine of a similar size.

\section{Conclusions}
The use case presented in this publication shows efficient facilitation of short-task computation for scientific workloads using serverless services in the digital twin domain, in particular in computational medicine. It is a promising approach for easy attainment of high parallelization for embarrassingly parallel jobs, and a viable alternative to HPC for all users who can access cloud resources. Our experiments prove that it is possible to acquire thousands of parallel sample computations and, as a result, speed up the calculation phase hundreds of times. We are certain that with an increase in limits, users will be able to achieve even better results. However, the range of supported models is restricted by service-specific considerations and there are various bottlenecks (network, CPU) that affect computational performance. Nevertheless, experiments show that communication overhead is a concern only for exceptionally quick tasks, becoming negligible with longer tasks. 

\section{Future work}
This particular implementation can be improved in many areas. One is to use multiplexing techniques (e.g. HTTP-pipelining or its successors in the newer versions of HTTP) whenever possible to minimize the number of open TCP connections. Another idea is to distribute the input data among many processes running the CloudVVUQ Executor, which could improve performance for IO-heavy workflows. As noted earlier, the possibility to collate an arbitrary number of inputs in a single request can accelerate the workflow for quick simulations by reducing overhead. Very promising results may be acquired by using  $ SQS \rightarrow Lambda \rightarrow DynamoDB $ architecture. Initial tests validate the approach and this architecture solved the issues of keeping alive TCP connections and using HTTP/1. Another noteworthy extension of our research could involve cost evaluation for serverless services as compared to HPC.

\section*{Acknowledgment}
This publication is supported by the European Union’s Horizon 2020 research and innovation programme under grant agreement Sano No 857533 and carried out within the International Research Agendas programme of the Foundation for Polish Science, co-financed by the European Union under the European Regional Development Fund.
This publication is also partly supported by the European Union’s Horizon 2020 research and innovation programme under grant agreement ISW No 101016503. \\
We thank Marek Konieczny from AGH for useful discussion and technical advice.

\IEEEtriggeratref{35}

\printbibliography

\end{document}